\documentclass[prb,twocolumn,showpacs,preprintnumbers,amsmath,amssymb,aps]{revtex4}

\usepackage{graphicx}

\newcommand{\be}{\begin{equation}}
\newcommand{\ee}{\end{equation}}
\newcommand{\bea}{\begin{eqnarray}}
\newcommand{\eea}{\end{eqnarray}}

\begin{document}


\title{Anisotropic magnetoresistance and piezoelectric effect in GaAs Hall samples}

\author{Orion Ciftja}
\affiliation{Department of Physics, Prairie View A\&M University, 
             Prairie View, Texas 77446, USA}

 \date{\today}

\begin{abstract}

Application of a strong magnetic field perpendicular to a two-dimensional electron system 
leads to a variety of quantum phases ranging from incompressible quantum Hall liquid
to Wigner solid, charge density wave and exotic non-Abelian states.
A few quantum phases seen on past experiments on GaAs Hall samples of electrons 
show pronounced anisotropic magneto-resistance values at certain weak magnetic fields. 
We argue that this might be due to the piezoelectric effect that is inherent in a semiconductor host like GaAs. 
Such an effect has the potential to create a sufficient in-plane internal strain that will be felt by
electrons and will determine the direction of high and low resistance. 
When Wigner solid, charge density wave and isotropic liquid phases 
are very close in energy, the overall stability of the system is very sensitive to local order and, 
thus, can be strongly influenced even by a weak perturbation 
such as the piezoelectric-induced effective electron-electron interaction which is anisotropic. 
In this work, we argue that an anisotropic interaction potential may stabilize 
anisotropic liquid phases of electrons even in a strong magnetic field regime 
where normally one expects to see only isotropic quantum Hall or isotropic Fermi liquid states. 
We use this approach to support a theoretical framework that envisions the possibility 
of an anisotropic liquid crystalline state of electrons in the lowest Landau level. 
%
%
In particular, we argue that an anisotropic liquid state of electrons may stabilize
in the lowest Landau level close to the liquid-solid transition region at filling factor $\nu=1/6$ 
for a given anisotropic Coulomb interaction potential. 
Quantum Monte Carlo simulations for a liquid crystalline state with broken rotational symmetry  
indicate stability of liquid crystalline order consistent with the existence of an anisotropic liquid state
of electrons stabilized by anisotropy at filling factor $\nu=1/6$ of the lowest Landau level.

\end{abstract}


\maketitle

\section{Introduction}
\label{sec-intro}

The study of properties of a strongly correlated two-dimensional 
electron system (2DES) has always been a fundamental goal of 
modern condensed matter physics. 
When a high mobility 2DES typically created in a host GaAs semiconductor environment 
is placed in a strong perpendicular magnetic field, 
purely quantum behavior dominates at very low temperatures. 
The combination of 2D confinement and magnetic field 
leads to a set of discrete, massively degenerate energy levels well 
separated from each other known as Landau levels (LLs)
labeled by an integer quantum number, $n=0, 1, \ldots$.
For simplicity of treatment one routinely assumes 
that the system of electrons is fully spin-polarized.
%
A key parameter that characterizes the properties of a 2DES 
in a perpendicular magnetic field is the filling factor, $\nu=N/N_s$ 
defined as the ratio of the number of electrons, $N$ 
to the degeneracy (number of available states), $N_s$ of each LL. 
For an integer number of completely filled LLs, we imagine 
that the electrons act independently and do not interact with 
each other~\cite{jmp-iqhe-one}. 
At very high magnetic fields, when electrons occupy only the 
lowest Landau level (LLL), correlation effects give rise to 
fractional quantum Hall effect (FQHE), 
a novel quantum many-body electronic liquid state~\cite{tsui82}.
The phase diagram of a 2DES in a strong perpendicular magnetic field
at filling factors $0 < \nu \leq 1$ is intricate with competing liquid and
Wigner solid phases.
At filling factors, $\nu=1/3, 2/5, 3/7, \ldots$ 
and $\nu=1/5, 2/9, \ldots$
electrons condense into an incompressible liquid FQHE state.
It is believed that at even-denominator filling factors, 
$\nu=1/2, 1/4$ and $1/6$ the electrons form 
a compressible Fermi liquid state~\cite{willett1,willett2}
while for filling factors, $ 0 < \nu \leq 1/7$, Wigner 
crystallization occurs~\cite{lam,esfarjani,zhu2}.
The principal FQHE states at filling factor $\nu=1/3$ and $1/5$ are thoroughly
explained and are well described by the 
Laughlin wave function approach~\cite{laughlin83}.
Other FQHE states at filling factors, $\nu=p/(2m p+1)$ (where $p$ and $m$ are integers) are 
readily understood in terms of the composite fermion (CF) 
theory~\cite{jain,physe}.
The $p \rightarrow \infty$ limit of such FQHE states~\cite{orepjb}
corresponds to even-denominator filled fractions, $\nu=1/(2m)$ 
that are believed to be compressible Fermi liquid states qualitatively 
different from the FQHE states of the originating sequence~\cite{hlr,rr,fantoni4}.
While the nature of various quantum states in the LLL seems to be well understood,
the situation is more complicated in higher LLs.
Fascinating cases are the anisotropic quantum Hall phases~\cite{lilly}
at filling factors $\nu=9/2, 11/2, \ldots$ in high LLs 
(where upper LLs with quantum numbers, $n \ge 2$ are partially filled).
The observed magneto-resistance anisotropy in such systems 
has been interpreted as strong evidence 
for the existence of a unidirectional (or striped) 
charge-density wave (CDW) state~\cite{fogler,moessner}.
A related possibility consistent with the experimental evidence would 
view the emergence of anisotropy in this regime as 
signature of a phase transition from an isotropic to an 
anisotropic electronic liquid crystalline phase~\cite{fradkin}. 
This way, one can view the onset of anisotropy as a transition to a quantum 
Hall nematic state described either as a 
Pomeranchuk-distorted phase~\cite{vadim,doan} or 
as a broken rotational symmetry (BRS) liquid 
crystalline phase~\cite{brshalf,ijmpb,orion2010,cite2011d}.
In all scenarios above, one assumes a Coulomb interaction between electrons.

\section{Microscopic Origin of Anisotropy }
\label{sec:origin}

%
The microscopic origin of the weak native rotational symmetry breaking
potential that may cause preferable orientation of electronic phases relative to 
the crystallographic axes of the host GaAs material
in a GaAs/AlGaAs hetero-structure, so far, has been unclear~\cite{cooper}. 
%
%
Studies in a tilted magnetic field~\cite{jungwirth,stanescu}
indicate that the strength of this anisotropic perturbation is only 
about $10^{-4} \, e^2/l_0 \approx 1 \, mK$  per electron
($l_0$ is the electron's magnetic length).
It is also expected that such an anisotropic interaction should be comparatively weak, 
because the effect of realignment of anisotropy in a tilted magnetic 
field has been observed~\cite{lilly2}.  
%
%
One possibility is that the GaAs crystalline substrate itself 
(coupled to electrons by exchange of phonons) induces a weak angular-dependent 
correction of the electron-electron Coulomb interaction~\cite{pan}. 
Emergence of a weak electron-phonon coupling 
(important below a certain low critical temperature) appears possible. 
%
%
%

Based on these considerations, it is reasonable to argue that 
the precise mechanism which breaks the in-plane symmetry of the 2DES 
originates from the GaAs host lattice and may be related to the crystal structure of GaAs
as well as the particular way of how electrons couple to the lattice 
in a given GaAs/AlGaAs hetero-structure.
We note that GaAs has a zinc-blende cubic structure which is described by two 
interpenetrating face-centered cubic (fcc) lattices. 
An important aspect of GaAs crystals is the absence of a center of 
inversion/symmetry. 
Hence, GaAs is capable of exhibiting piezoelectric and related effects 
depending on polar symmetry. Usually the piezoelectric interaction in 
semiconductors is not of major importance. 
However, in high quality crystals (such as GaAs Hall samples) 
and in very low temperatures, this interaction, while weak, can play a role. 
Given that GaAs is a piezoelectric crystal, we remark that 
it is possible that 
the piezoelectric interaction, which is anisotropic for cubic crystals, 
plays a role in the anisotropic orientation of the given electronic structure
hosted there.
Clearly, it is possible that the piezoelectric effect 
inherent in GaAs semiconductor samples can create a sufficient in-plane internal strain 
which determines the direction of high and low resistance in GaAs Hall samples
and, thus, may induce some native anisotropy in the hosted system of electrons.
%
%

At this juncture, the key point made is that an anisotropic 
perturbation term (in general) may explain some 
basic physics features that control the eventual emergence 
of anisotropic liquid states of electrons in the quantum Hall regime.
As such, this anisotropic term might play an important role in the stability of 
various electronic phases in all LL-s, including the LLL.
In particular,  when Wigner solid, CDW and isotropic liquid phases are very close in energy, 
the overall stability of the system is very sensitive to local order and, thus, 
can be strongly influenced even by an anisotropic perturbation
like the piezoelectric-induced one.
This argument leads one to believe that anisotropic phases of electrons may arise 
even in the strong magnetic field regime (in the LLL) at settings 
where one would normally expect isotropic fractional quantum Hall liquid, 
isotropic Fermi liquid, or Wigner solid states. 
This mechanism may be appealed to explain recent experiments that indicate the presence of 
a new anisotropic fractional quantum Hall effect state of 
electrons at regimes not anticipated before~\cite{abstract3}. 
%

\section{Anisotropic phases in the lowest Landau level}
\label{brslll}

In earlier work~\cite{brsall,hexatic}, 
we investigated the possible existence of anisotropic liquid phases
of electrons in the LLL at Laughlin fractions, $\nu=1/3$, $1/5$ and $1/7$.
We introduced many-body trial wave functions that are translationally invariant
but possess twofold, fourfold, or sixfold BRS at respective filling factors,
$\nu=1/3$, $1/5$, and $1/7$ of either the LLL, or excited LLs.
We considered a standard Coulomb interaction potential between electrons
as well as two other model potentials that include thickness effects.
All the electron-electron interaction potentials considered in such a study 
were isotropic~\cite{brsall}.
One of the specific findings was that all the anisotropic liquid crystalline phases considered 
had higher energy than the competing isotropic Laughlin liquid states in the LLL 
for systems electrons interacting with the usual isotropic Coulomb potential. 
%
Based on these findings we concluded that,
if there are electronic liquid states in the LLL, 
these states are isotropic and should possess rotation symmetry.

%
%

However,  presence of an anisotropic interaction term in the Hamiltonian 
reopens the problem and suggests a re-examination of the possibility 
of anisotropic liquid phases of electrons at all filling factors in the LLL 
where isotropic liquid states were previously known to exist.
%
%
%
%
%
%
We have identified the very fragile isotropic 
Fermi liquid state~\cite{kun-yang} at $\nu=1/6$ 
(just before the onset of Wigner crystallization) 
as a good candidate that may be destabilized 
by an electron-electron anisotropic perturbation. 
With this idea in mind, we looked at $\nu=1/6$ and first compared the 
stability conditions of various known phases 
such as Wigner solids, Fermi liquids, Bose Laughlin liquids
(Wigner crystallization occurs around $\nu=1/6.5$).
We found out that different quantum states, 
for instance, a CF Fermi liquid and a Bose Laughlin state~\cite{bose} 
have energies extremely close to each other with differences
in the order of  $10^{-4} \, e^2/l_0$. 
%
Hence, an anisotropic perturbation of comparable magnitude
(that can readily be induced)  
may suffice to shift the energy balance in favour of an anisotropic liquid phase. 
%
%
%
%
%
%
%
%

%
%
%
%
%
%
\begin{table*}[ht!]
\caption[]{
           Roman-numbered energy shells (in $\vec{k}$-space) in increasing order of energy
           where 
           $g$ is the degeneracy of each energy value, 
           $N$ is the total number of electrons (for a spin-polarized system) and 
           $(n_x,n_y)$ are the corresponding plane wave quantum numbers. }
\label{tabkspace}
\begin{center}
\begin{tabular}{|c|c|c|c|c|}
\hline                                                  
Shell & $n_x^2+n_y^2$  & $g$  & $N$  & $(n_x,n_y)$          \\ \hline
I     & 0                & 1    & 1    & $(0,0)$              \\ \hline
II    & 1                & 4    & 5     & $(-1,0),(0,-1),(0,1),(1,0)$  
                                                          \\ \hline
III   & 2                & 4    & 9   & $(-1,-1),(-1,1),(1,-1),(1,1)$ 
                                                          \\ \hline
IV    & 4                & 4    & 13   & $(-2,0),(0,-2),(0,2),(2,0)$
                                                          \\ \hline
V     & 5                & 8    & 21   & $(-2,-1),(-2,1),(-1,-2),(-1,2),
                                    (1,-2),(1,2),(2,-1),(2,1)$
                                                          \\ \hline
VI    & 8               & 4     & 25  & $(-2,-2),(-2,2),(2,-2),(2,2)$     
                                                          \\ \hline
VII   & 9               & 4     & 29  & $(-3,0),(0,-3),(0,3),(3,0)$ 
                                                          \\ \hline
VIII  & 10                & 8   & 37    & $(-3,-1),(-3,1),(-1,-3),(-1,3),
                                     (1,-3),(1,3),(3,-1),(3,1)$
                                                          \\ \hline
IX    & 13              & 8     & 45  & $(-3,-2),(-3,2),(-2,-3),(-2,3),
                                     (2,-3),(2,3),(3,-2),(3,2)$
                                                          \\ \hline
X     & 16              & 4     & 49  & $(-4,0),(0,-4),(0,4),(4,0)$ 
                                                          \\ \hline
XI    & 17                & 8   & 57    & $(-4,-1),(-4,1),(-1,-4),(-1,4),
                                     (1,-4),(1,4),(4,-1),(4,1)$
                                                          \\ \hline
\end{tabular}
\end{center}
\end{table*}
%
%
%
%


Typically, the Hamiltonian for a 2DES of $N$ electrons of isotropic mass, $m_e$
and charge, $-e (e>0)$ in a perpendicular magnetic field,
$\vec{B}=(0, 0, -B)= \vec{\nabla} \times \vec{A}(\vec{r})$ (symmetric gauge)
is written as:
\begin{equation}
\hat{H}=\hat{K}+\hat{V} \ ,
\label{hamiltonian}
\end{equation}
where 
$\hat{K}=\frac{1}{2 \, m_e} \sum_{i=1}^{N}
\left [ \hat{ \vec{p}}_i +e \, \vec{A}(\vec{r}_i) \right]^2$
%
%
is the kinetic energy operator 
%
%
and $\hat{V}$ is the potential energy operator:
\begin{equation}
\hat{V}=\hat{V}_{ee}+\hat{V}_{eb}+\hat{V}_{bb} \ , 
\label{pot_en}
\end{equation}
where $\hat{V}_{ee}, \hat{V}_{eb}$ and $\hat{V}_{bb}$ 
are the electron-electron (ee), electron-background (eb) and
background-background (bb) potential energy operators.
The interaction potential is usually assumed to be
the (isotropic) Coulomb potential,
$v_{C}(r_{ij})=e^2/r_{ij}$,
where $r_{ij}=|\vec{r}_i-\vec{r}_j|$ 
is the distance between two electrons
(as customary, Coulomb's electric constant is not included in the expressions
of the interaction potential and/or energy).
A standard model for a 2DES in a disk geometry assumes that $N$ fully 
spin-polarized electrons are embedded in a uniform  neutralizing
background disk of area $\Omega_N=\pi \, R_{N}^2$ and radius $R_N$.
The electrons can move freely all over the 2D space and are not constrained
to stay inside the disk.
The uniform density of the system can be written as 
$\rho_0=\nu/(2 \pi l_0^2)$
where $l_0=\sqrt{\hbar/(e \, B)}$ is the magnetic length.
The radius of the disk is determined from the condition:
$\rho_0=N/\Omega_{N}$.
%
%
%
%
%
%

%
%
%
%

To account for the influence of the GaAs crystalline substrate (for instance, 
through the piezoelectric effect) we will have to
ammend the given Hamiltonian with an anisotropic 
electron-electron perturbation term:
\begin{equation}                     
v(r_{ij},\theta_{ij})=v_C(r_{ij})+\Delta v(r_{ij},\theta_{ij}) \ ,
\label{perturb}
\end{equation}                     
where $v_C(r_{ij})$  is the bare Coulomb potential 
and   
$\Delta v(r_{ij},\theta_{ij})$
is an additional substrate-related anisotropic perturbation term.
Here $\theta_{ij}$ 
is some angle between the radial vector, $\vec{r}_{ij}=( x_{ij}, y_{ij} )$ 
between electron's positions and the crystallographic axes of the GaAs substrate. 
%
%
%
%
%
%
In absence of a realistic ab-initio potential for $\Delta v(r_{ij},\theta_{ij})$, 
we choose to adopt a phenomenological approach and
considered an anisotropic Coulomb interaction potential of the form:   
\begin{equation}
v_{\gamma}( x_{ij}, y_{ij}) = 
\frac{e^2}{\sqrt{x_{ij}^2/\gamma^2+\gamma^2 \, y_{ij}^2}} \ ,
\label{distorted}
\end{equation}
where the interaction anisotropy parameter, $\gamma \ge 1$ 
tunes the degree of anisotropy of  the electron-electron interaction potential. 
Note that $v_{\gamma=1}(x_{ij}, y_{ij})=v_{C}(r_{ij})$
where $r_{ij}=\sqrt{ x_{ij}^2+y_{ij}^2}$.
This ammendment affects only the $\hat{V}_{ee}$ operator in Eq.(\ref{pot_en}).
%
%
%
%
In order to build a suitable microscopic wave function with anisotropic 
features, we started with a BRS wave function for filling factor $\nu=1/2$ that we 
had introduced in an earlier work~\cite{brshalf} and generalized it 
to filling factor $\nu=1/6$.
This approach lead us to an anisotropic BRS liquid crystalline state appropriate for 
a system of electrons at filling factor $\nu=1/6$ that was written in the following form:
%
%
%
%
%
\begin{eqnarray}
 \Psi_{\alpha} &=& 
\prod_{i>j}^{N} \, (z_i-z_j)^4
 (z_i-z_j+\alpha)(z_i-z_j-\alpha)  
\nonumber \\ & &
\times  \exp{\left( -\sum_{j=1}^{N} \frac{|z_j|^2}{4 \, l_0^2} \right)}
           det \left | e^{i \, \vec{k}_{\alpha} \, \vec{r}_j} \right |  \ ,
\label{bosealpha}
\end{eqnarray}
where $N$ is the number of electrons that occupy the $N$ lowest-lying 
plane wave states labeled by the momenta $\{ \vec{k}_{\alpha} \}$ 
of an ideal 2D spin-polarized Fermi gas,
$z_j=x_j+i \, y_j$ is the 2D position coordinate in complex notation
and we discard the LLL projection operator
(with the usual assumption that kinetic energy is not very important).
%
%
The allowed plane wave states for an ideal 2D spin-polarized Fermi gas~\cite{2013d}
have an energy of
$E(\vec{k})=\frac{\hbar^2}{2 \, m_e} \, | \vec{k} |^2$
where $\vec{k}=\frac{2 \, \pi}{L} \, \vec{n}$ and $\vec{n}=(n_x,n_y)$
with $n_{x,y}=0, \pm 1, \pm 2, \ldots$.
The value of $L$ is such that $L^2=\pi \, R_{N}^2$ for any given $N$.
In our case we considered systems with a number, $N$ of electrons
that corresponds to closed shells (in $\vec{k}$-space).
%
%
Different $\vec{k}$-states may have the same energy value, $E(\vec{k})$ 
and, thus, they belong to the same energy shell.
The number of different $\vec{k}$-states for a given energy shell represents
the degeneracy of that energy value.
In Table.~\ref{tabkspace} we list such energy shells in increasing order of
energy (roman-numbered) with their degeneracies, 
corresponding quantum states, $(n_x,n_y)$ 
and the resulting total number of electrons, $N$ for closed shells.
This means that our simulations were performed with systems 
containing $N=5, 9, \ldots, 57$ electrons.

The (generally complex) anisotropy parameter, $\alpha$ of the anisotropic BRS liquid wave function 
can be considered as a nematic director whose phase is associated with
the angle relative to GaAs hard resistance crystalline axis.
Note that the wave function in Eq.(\ref{bosealpha}) reduces
to a generalized version of the Rezayi-Read (RR) wave function~\cite{rr}
for $\nu=1/6$ when $\alpha=0$.
In our case, we consider $\alpha$ to be real so that the 
system has a stronger modulation in  the $x$-direction.
%
%
%
The wave function, $\Psi_{\alpha}$ is antisymmetric and translationally invariant,
but lacks rotational symmetry when $\alpha \neq 0$.
Thus, it is an obvious starting point for a {\em nematic} 
anisotropic liquid state at $\nu=1/6$.
%
%
The (minimal) splitting of the zeroes in the Laughlin polynomial part of the wave function
follows the same rationale as that of the $\nu=1/3$ BRS Laughlin state~\cite{musaelian}.
%

%
%
%
It is believed that the $\alpha=0$ RR Fermi liquid wave function in Eq.(\ref{bosealpha}) 
lies quite close to the true ground state of the system at filling factor $\nu=1/6$ 
for the case of electrons interacting with an isotropic Coulomb interaction potential.
However, this filling factor also represents the approximate liquid-solid crossing point.
Thus, other competing phases~\cite{jltp2016}, 
for instance crystal states, may intervene to destabilize it.
One class of competing states that has been recently studied in detail 
is a CF crystal state for filling factors between $\nu=1/5$ and $2/9$ in the LLL~\cite{archer}.
Although the range of filling factors from $\nu=1/5$ to $2/9$ is slightly larger than 
filling factor $\nu=1/6$ considered in this work,
one gets the feeling that the competing energies involved 
(CF crystal state versus liquid state) are very close to each other.
It is worthwhile mentioning that the CF crystal results mentioned above 
deal with a system of $N=96$ electrons in a spherical geometry~\cite{archer}. 
A direct comparison of the energy results between the isotropic $\alpha=0$ RR
Fermi liquid state and the CF crystal state energies~\cite{archer} 
would be difficult at this juncture.
This is because of differences in model, geometry and treatment that exist 
between the current study in disk geometry and its counterpart in spherical geometry
(for a given finite system of $N$ particles, the energy per particle in a disk geometry 
 is different from its counterpart in a spherical geometry, 
LLL projection causes differences, and so on.).

\begin{figure}[!t]
\begin{center}
\includegraphics[width=3.4in]{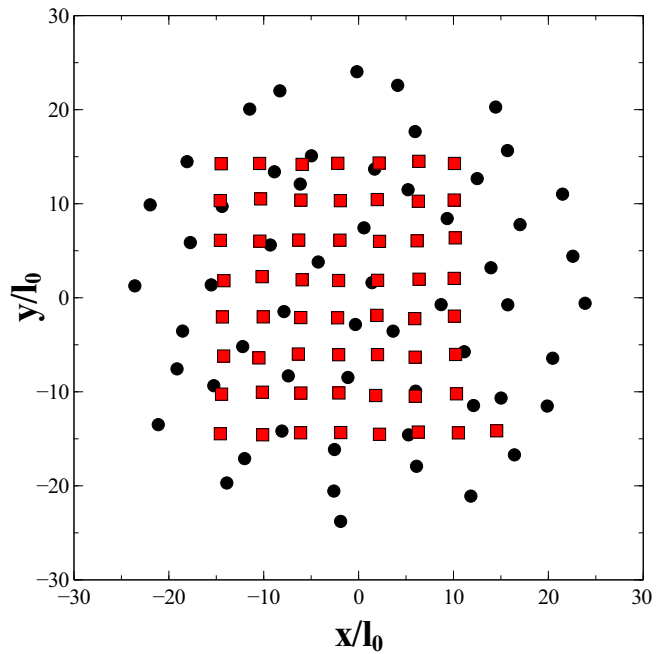}
\end{center}
\caption[]{ 
             Snapshot of the final configuration of a system of 
                 $N=57$ electrons (filled circles)
             after a full QMC run (disk geometry). 
             The system of electrons is described by an 
             anisotropic Fermi liquid BRS wave function with $\alpha=5$ 
             and represents a state with filling factor $\nu=1/6$ of the LLL.
             The initial configuration of the system at the start of the QMC run is also 
             given (filled squares). 
             Note that the electrons are initially placed in a regular square lattice
             and, then, are randomly displaced by a little bit relative to the lattice
             sites for better randomization before the start of a QMC run. 
            }
\label{fig1}
\end{figure}
%
%
%
%

%
%
In our study we resort to the approximation of neglecting the LLL projection operator 
(LLL projection would affect the plane wave states of the Slater determinant).
The kinetic energy would have been a mere constant if the whole wave function 
would have been properly projected~\cite{kamilla}.
Although lack of LLL projection may affect quantitatively the energy values for a given
system of electrons, it will not change qualitatively any of the main conclusions 
drawn since we are interested in energy differences.
With other words, energy differences will likely be quite accurate (with or without LLL projection)
since separately the energies are calculated within the same approximation.
Hence, the discrepancies/errors which are systematic will be of the same 
order of magnitude and would tend to cancel out when energy differences are calculated.
%
%
Earlier studies of isotropic CF wave functions (constructed by multiplying the
wave functions for filled LLs with Jastrow-Laughlin correlation factors)
have shown that such correlation factors ensure that the resulting states 
lie predominantly in the LLL~\cite{trivedi}. 
%
%
%
%
For simplicity of notation, let's denote as $P_{N}(\nu)=\prod_{i>j}^{N}(z_i-z_j)^{1/\nu}$ 
the Laughlin polynomial factor of an isotropic state at filling factor, $\nu$. 
For instance, $P_{N}(\nu=1/3)=\prod_{i>j}^{N}(z_i-z_j)^{3}$ 
represents the typical isotropic polynomial factor of a Laughlin wave function 
at filling factor $\nu=1/3$ of the LLL.
Note that the $\alpha$-dependent anisotropic polynomial factor in the BRS wave function 
of Eq.(\ref{bosealpha}) can be viewed as: 
$P_{N}(\nu=1/6)-\alpha^2 \, P_N(\nu=1/4)$. 
As already pointed out, each of the isotropic unprojected wave functions, separately, 
are expected to have tiny amplitudes in higher LLs. 
Since the anisotropic BRS wave function can be viewed as a linear combination of two isotropic counterparts, 
the conclusion reached for each isotropic portion is expected to apply to the linear combination.
This means that the presence of Jastrow-Laughlin polynomial factors 
with relative high powers in Eq.(\ref{bosealpha})
already provides a considerably projection into the LLL~\cite{trivedi}
reassuring us on the validity of the approach.

%
%
%
%
\begin{figure}[!t]
\begin{center}
\includegraphics[width=3.4in]{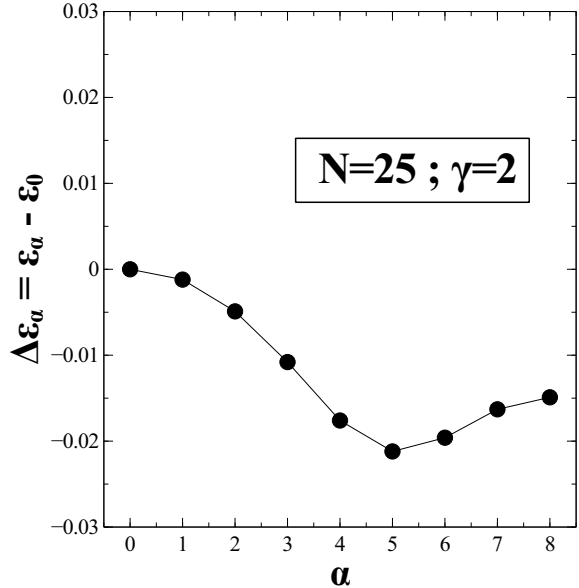}
\end{center}
\caption[]{ 
            Difference of energy (per electron), 
           $\Delta \epsilon_{\alpha}=\epsilon_{\alpha}-\epsilon_{0}$ as a function of the 
            anisotropy parameter, $\alpha$. 
            The energy, $\epsilon_{\alpha}$ is the energy (per electron) of the
            anisotropic BRS liquid state while, $\epsilon_{0}$ is its isotropic liquid counterpart.
            Results are obtained after QMC simulations in a disk geometry for a system 
            of $N=25$ electrons at filling factor $\nu=1/6$ of the LLL.
            Electrons interact with an anisotropic Coulomb interaction potential, 
            $v_{\gamma=2}(x,y)$
            where $\gamma=2$ represents the value of the interaction anisotropy parameter.
            Energies are in units of $e^2/l_0$.  
            }
\label{fig2}
\end{figure}
\begin{figure}[!t]
\begin{center}
\includegraphics[width=3.4in]{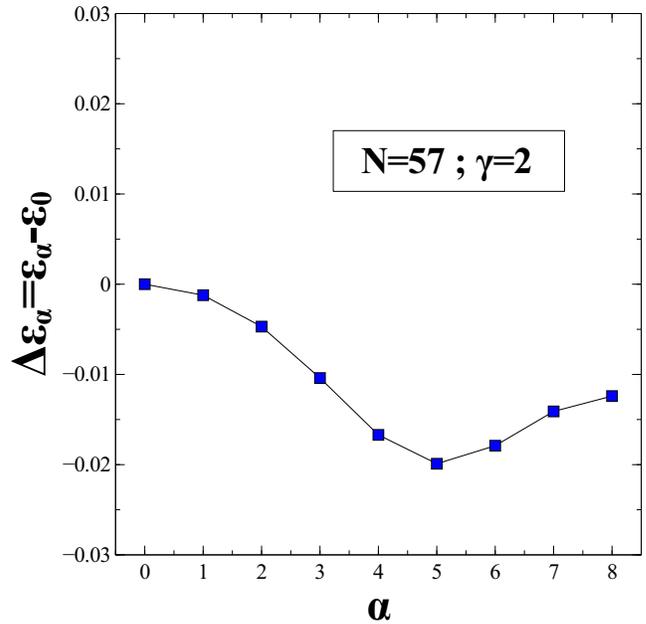}
\end{center}
\caption[]{ Same as in Fig.~\ref{fig2} but for a larger system of $N=57$ electrons.
                 The energy differences for the larger system of $N=57$ electrons are
                  very close to values seen for the smaller system of $N=25$ electrons. 
            }
\label{fig-dener-g2-n57}
\end{figure}
%
%
%

%
%
%
It has been recently pointed out that the topological
description of FQHE wave functions is not complete since
the usual implicit assumption of rotational symmetry
hides key  geometrical features~\cite{haldane2011}.
Along these lines,
one must distinguish between several "metrics" that 
naturally arise in a real electronic system.
One metric derives from the  band mass tensor of electrons. 
A second metric is controlled by the dielectric properties 
of the semiconducting material that hosts the electrons~\cite{haldane2011}.
The second metric determines the nature of the interaction 
between electrons.
The rotational invariance of the system is lifted if 
these two metrics are different from one another.
What counts is the relative difference between them~\cite{bo}.
Thus, one can, for simplicity, assume an isotropic mass tensor and 
anisotropic dielectric tensor.
In other words, as done in this work, one can assume that 
the mass of electrons is isotropic, 
but treat the interaction between electrons as anisotropic.
It turns out that, despite differences in specific details, 
the discussion in Sec.~\ref{sec:origin},
the model/treatment and the wave function considered here 
dovetail and are fully consistent with the spirit of Haldane's ideas~\cite{haldane2011}.

\section{Results and conclusions}
\label{sec-summary}

%
%
%
%
\begin{figure}[!t]
\includegraphics[width=3.4in]{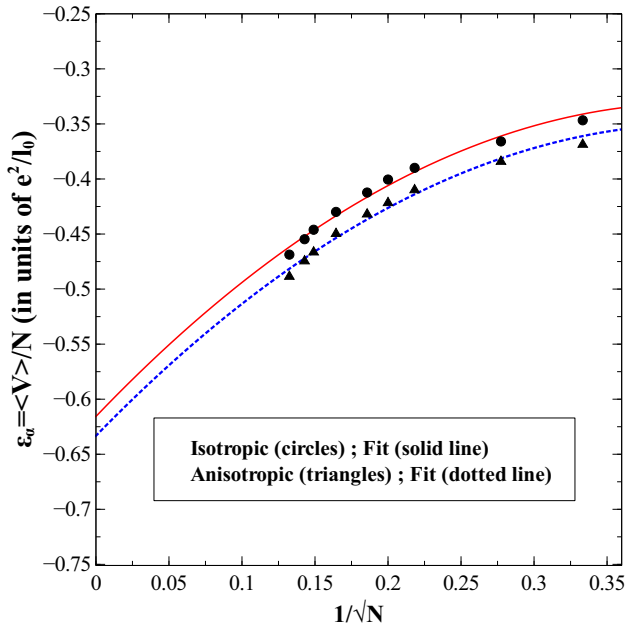}
\caption{  \label{fig:fit-g2-n}
        Monte Carlo results in disk geometry corresponding to an isotropic ($\alpha=0$) and
        anisotropic ($\alpha \neq 0$) Fermi liquid states at filling factor $\nu=1/6$ of the LLL.
        The interaction energy per electron, $\epsilon_{\alpha}=\langle \hat{V} \rangle/N$, 
        is plotted as a function of $1/\sqrt{N}$ for systems with 
        $N=5, 9, 13, 21, 25, 29, 37, 49$ and $57$ electrons.
        Filled circles correspond to energies of isotropic states ($\alpha=0$) 
        where the solid line is the least square fit function in Eq.(\ref{fit_isotrop}).
        Filled triangles correspond to energies of anisotropic states 
        (at optimal $\alpha_{0} \neq 0$) 
        where the dotted line is the least square fit of Eq.(\ref{fit_anisotrop}).  
        Electrons interact with an anisotropic Coulomb interaction potential, 
        $v_{\gamma=2}(x,y)$.
        Statistical uncertainty is of the size of the symbols used.
        Energies are in units of $e^2/l_0$.
}
\end{figure}
%
%

%
%
%
%
%
\begin{figure}[!t]
\includegraphics[width=3.4in]{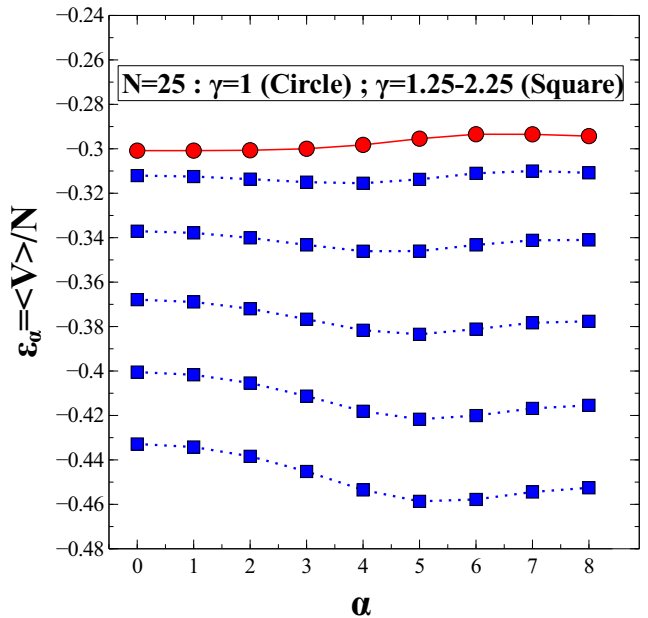}
\caption{  \label{fig:ener_n25_g}
        Energy per electron as a function of wave function anisotropy parameter, $\alpha$
        for an anisotropic Coulomb interaction potential, $v_{\gamma}(x,y)$
        with interaction anisotropy parameter 
        values varying from $\gamma=1$ (standard isotropic Coulomb interaction potential) 
        up to $\gamma=2.25$ 
        (largest strength of the  anisotropic Coulomb interaction potential considered).
        Filled circles correspond to $\gamma=1$ (isotropic Coulomb).
        Filled squares correspond to $\gamma=1.25, 1.50, 1.75, 2.00$ and $2.25$ 
        (from top to bottom).
        The system under consideration consists of $N=25$ electrons in disk geometry 
        at filling factor $\nu=1/6$ of the LLL.
        The value, $\alpha=0$  represents an isotropic Fermi liquid phase.
        Any other value,   $\alpha > 0$ represents anisotropic BRS Fermi liquid states.
        Statistical uncertainty is of the size of the symbols used.
        Solid/dotted lines are guides for the eye. 
        Energies are in units of $e^2/l_0$.
}
\end{figure}
%
%
%

In our calculations,
we consider small finite systems of electrons at filling factor $\nu=1/6$ of the LLL.
The electrons are confined in a uniformly charged background disk 
and interact with each other via the anisotropic Coulomb interaction potential,  $v_{\gamma}(x,y)$
of Eq.(\ref{distorted}).
The system of electrons under consideration is described by the 
BRS Fermi liquid wave function in Eq.(\ref{bosealpha}).
We performed detailed quantum Monte Carlo (QMC) 
simulations~\cite{mcdisk} in disk geometry 
in order to assess the stability of given BRS liquid crystalline phases
relative to the isotropic liquid counterpart. 
QMC simulations enabled us to calculate accurately the potential energy (per electron), 
$\epsilon_{\alpha}=\langle \hat{V} \rangle/N$
for a chosen set of values of the anisotropy parameter
ranging from $\alpha=0$ (isotropic) to $\alpha=8$ (the largest value considered)
in steps of $1$.
From now on, it is implied that values of $\alpha$ are given in
units of the magnetic length, $l_{0}$.
We, initially, choose an anisotropic Coulomb interaction potential with $\gamma=2$
to start the calculations.
%
%
We considered various systems with $5 \leq N \leq 57$ electrons.
As can be seen from Fig.~\ref{fig1} the distribution of electrons 
somehow reflects the built-in anisotropy of the BRS liquid crystalline wave function.
%
After obtaining the energies of all systems considered,
we calculated the energy difference between anisotropic BRS liquid crystalline
states and their isotropic liquid counterpart:
%
%
$\Delta \epsilon_{\alpha}=\epsilon_{\alpha}-\epsilon_{0}$ 
as a function of the anisotropy parameter, $\alpha$.
%
%
%
%
In Fig.~\ref{fig2} we show the results for $\Delta \epsilon_{\alpha}$
for a system of $N=25$ electrons where electrons interact with the
anisotropic Coulomb interaction potential, $v_{\gamma=2}(x,y)$. 
The statistical uncertainity of the results is smaller than the size
of the symbols (note that energy differences are generally very accurate).  
It is evident that the anisotropic BRS liquid state of electrons 
is energetically favored at all instances.
Even though we did not try a full optimization of the energy as a function
of the continuous parameter, $\alpha$, the QMC simulations clearly indicated that 
there is always a value $\alpha \neq 0$ 
(namely, there is an anisotropic liquid state of electrons)
that has lower energy than the isotropic one.
For the given choice of discrete, $\alpha=0, 1,  \ldots, 8$ values,
the lowest energy of a system of $N=25$ electrons 
was achieved for an optimal value of $\alpha_0=5$.
For the case study of $N=25$ electrons interacting with an
anisotropic Coulomb interaction, $v_{\gamma=2}(x,y)$, 
the amount of gain in energy was found to be approximately
$\sim 20 \times 10^{-3} \, {e^2}/{l_0}$ which is
quite sizeable in a quantum Hall energy scale.
%
%
As can be seen from Fig.~\ref{fig-dener-g2-n57},
simulations with a larger number of $N=57$ electrons 
and same $v_{\gamma=2}(x,y)$ anisotropic interaction potential
show similar qualitative features as those for $N=25$ electrons.
This means that energy differences are almost size-independent.

%

%
%
%
%
%
%
In order to obtain a reasonable bulk estimate of the energy of the system 
in the thermodynamic limit
for the $v_{\gamma=2}(x,y)$ anisotropic interaction potential,
we performed a careful finite-size analysis of the available data 
for systems of $5 \leq N \leq 57$ electrons. 
We followed wellknown procedures~\cite{morfhalperin}
to fit the available energies with a quadratic polynomial function (of $1/\sqrt{N}$).
This was done for both isotropic Fermi liquid states ($\alpha=0$) 
and anisotropic BRS liquid states (those with the lowest energy for an optimal 
value, $\alpha_0 \neq 0$).
For the isotropic case ($\alpha=0$), we found:
\begin{equation}
\epsilon_0=
    \left( -0.61565  +\frac{1.38519}{\sqrt{N}} -\frac{1.68438}{N}  \right)
    \frac{e^2}{l_0}  \ .
\label{fit_isotrop}
\end{equation}
The result of the fit for the optimal anisotropic BRS energies was:
\begin{equation}
\epsilon_{\alpha_0}=
    \left( -0.63322 +\frac{1.36197}{\sqrt{N}} -\frac{1.63615}{N}  \right)
    \frac{e^2}{l_0}  \ .
\label{fit_anisotrop}
\end{equation}
In Fig.~\ref{fig:fit-g2-n} we show the interaction energy per electron
plotted as a function of $1/\sqrt{N}$ together with the results of the fit.
Extrapolation of the results (for $N \rightarrow \infty$) provides a
useful estimate to the energy  
in the thermodynamic limit (the first term in each of the parentheses). 
%
%
Although the convergence of the results (as a function of $N$) is slow (very typical)
and we are limited by our computational power to systems with up to $N \leq 57$ electrons, 
the results of the extrapolation in the $N \rightarrow \infty$ limit seem unambiguous
in suggesting  that the lower energy of the anisotropic BRS Fermi liquid
state persists in the bulk limit.
%
%
%
%

%
%
%
%
%
The value of the interaction anisotropy parameter, $\gamma$ is set phenomenologically.
We choose to set it initially to $\gamma=2$ and the 
results in Fig.~\ref{fig2}, Fig.~\ref{fig-dener-g2-n57} and Fig.~\ref{fig:fit-g2-n} reflect this choice.
This choice of $\gamma$, while not special, seems to be somewhere in the middle
of the range of choices seen on related studies dealing with a similar topic.
%
%
For instance, a recent work on the effect of the same anisotropic Coulomb interaction potential
on the $\nu=1/3$ FQHE state employing an exact diagonalization method 
for $N=10$ electrons in torus geometry~\cite{haowang} uses values of
interaction anisotropy parameter that range from $1$ (isotropic) 
to way larger than $4$ or $5$ (anisotropic).
That said, a legitimate question that arises is whether the same behavior as seen 
for $\gamma=2$ persists if we use smaller values of the 
interaction anisotropy parameter, $\gamma$.
To check the situation, we performed additional simulations for a finite system of $N=25$ 
electrons at filling factor $\nu=1/6$ of the LLL using different $\gamma$ values
that are smaller than the initially considered value of $\gamma=2$.
This way we can see whether the isotropic Fermi liquid state survives the 
breakdown of rotational invariance for smaller interaction anisotropy values.
To this effect we considered a series of values for the interaction anisotropy parameter, $\gamma$ 
that vary from $1$ (isotropic Coulomb interaction potential) to $2.25$ and in between in steps of $0.25$.
%
%
The results in Fig.~\ref{fig:ener_n25_g} for $\gamma=1, 1.25, 1.50, \ldots$ 
and systems of $N=25$ electrons
clearly show that any value $\gamma \neq 1$ considered 
leads to an anisotropic BRS liquid phase at filling $\nu=1/6$,
while the isotropic liquid state is always stable for $\gamma=1$. 
The results seem to suggest that any value, $\gamma > \gamma_c$ 
where $\gamma_c$ is between $1$ and $1.25$ (and possibly quite close to $1$)
will lead to the stabilization of an anisotropic BRS liquid state.
However, a precise determination of $\gamma_c$  would be hard to quantify 
based on the numerical accuracy limitations of these calculations 
when calculating the expected very small energy differences between states in such limit. 
%
%
Since the parameter, $\alpha$ in the BRS wave function, $\Psi_{\alpha}$ is adjustable, 
the present calculations test the variational principle for a Hamiltonian with isotropic ($\gamma=1$) and 
anisotropic ($\gamma>1$) interaction. 
In both cases, the energy (namely, the expectation value of the quantum Hamiltonian 
with respect to the chosen wave function) develops a minimum. 
The minimum occurs at $\alpha=0$ for the isotropic ($\gamma=1$) interaction and 
at a nonzero value, $\alpha_0 \neq 0$ for the anisotropic ($\gamma>1$) case. 
Therefore, a related question that one might want to ask is how $\alpha_0$ varies as $\gamma \rightarrow 1$.
Hence, a study of the $\alpha$-$\gamma$ relationship 
in the $\gamma \rightarrow 1$ limit is of great interest.
When $\gamma=1$, the value $\alpha=0$ is the natural minimum for the energy of this family of wave functions. 
This fact can be used to obtain a qualitative understanding of how the overlap 
between the $\alpha \neq 0$ and the $\alpha=0$ state varies 
since one can see from Eq.(\ref{bosealpha}) that:
\begin{equation}
\Psi_{\alpha}= \Biggl[ 1 - \alpha^2 \prod_{i>j}^{N} \frac{1}{ (z_i-z_j)^2} \Biggr] \, \Psi_{\alpha=0} \ .
\label{overlap}
\end{equation}
Apart normalization factors, one notices that
$\langle \Psi_{\alpha=0} | \Psi_{\alpha} \rangle \approx 1-c \, \alpha^2$, 
where $c$ represents a constant.

%
%

%
%
When $\alpha=0$, the wave function in Eq.(\ref{bosealpha})
represents an isotropic Fermi liquid state.
It appears clear that, for this set of wave functions,
the $\alpha=0$ BRS wave function in Eq.(\ref{bosealpha}) 
represents the mininum energy state when electrons interact with 
a standard isotropic Coulomb interaction potential ($\gamma=1$).
For a system of $N=25$ electrons, 
we found that $\epsilon_{\alpha=0}(\gamma=1) = -0.30083 \, e^2/l_0$ 
(see Fig.~\ref{fig:ener_n25_g}).
This energy value compares very favorably with the 
expected CF Wigner crystal energy at the same filling factor.
The study in Ref.~\citenum{archer} provides a useful reference energy 
(that serves as a broad guideline of typical ground 
state energies as a function of filling factor).
It is denoted as $E_{fit}$ in the caption of Fig.~2 of Ref.~\citenum{archer}
(a figure that reflects results obtained with systems of $N=96$ electrons in spherical geometry).
%
%
We used the expression for $E_{fit}$ that was provided to obtain an estimate
of the expected CF crystal state energy at filling factor $\nu=1/6$
resulting in
$E_{fit}(\nu=1/6)= -0.299477 \, e^2/l_0$.
A crude analysis of the data available, indicates further decrease
of the $\epsilon_{\alpha=0}(\gamma=1)$ energy when $N$ increases.
This trend is typical for systems of electrons 
in a disk geometry (for instance, see Fig.~\ref{fig:fit-g2-n}).
The indication is that the $\alpha=0$ isotropic Fermi liquid state energies
for the $\nu=1/6$ state of electrons interacting with a standard Coulomb interaction potential
are close and lower than the $E_{fit}(\nu=1/6)$ value.
%
However, as we already cautioned earlier in Sec.~\ref{brslll},
one should not draw any definitive conclusions
since in this case we are comparing results that are obtained under 
different conditions, namely disk versus spherical geometry and LLL unprojected versus 
projected wave function (the case of the CF Wigner crystal states).
What is definitely clear from our results is a substantial decrease of the energy achieved 
by the BRS wave function in Eq.(\ref{bosealpha})
for the case of the anisotropic $v_{\gamma=2}(x,y)$ interaction potential.
%
%
A useful hint that we can draw from this finding is to suggest 
that a "deformed/distorted" (via the splitting of the zeroes mechanism)
CF Wigner crystal state may be a better starting point to study Wigner crystal phases 
in the LLL for a $v_{\gamma>1}(x,y)$ 
anisotropic interaction potential between electrons as the one considered here.

%
%
%
%
\begin{figure}[!t]
\includegraphics[width=3.4in]{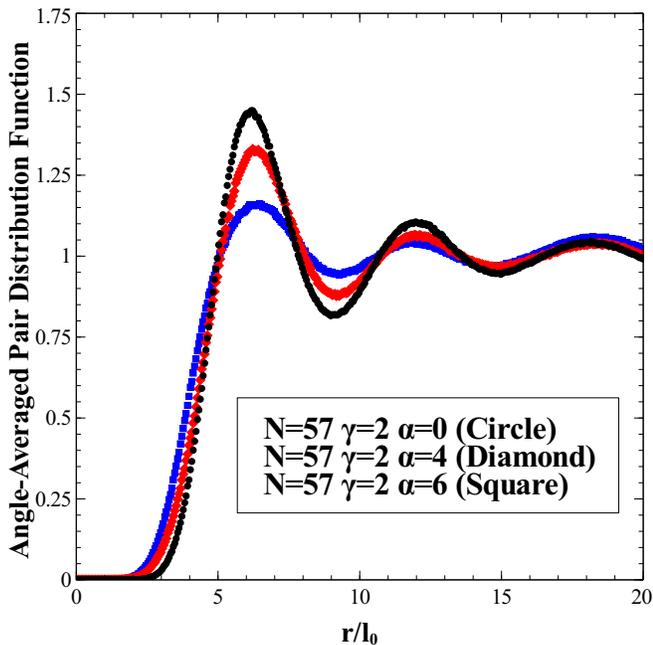}
\caption{  \label{fig:gn57}
        Angle-averaged pair distribution function, $g(r)$ as a function
        of dimensionless distance, $r/l_0$ for a system of $N=57$ electrons.
       The system is described by an anisotropic BRS Fermi liquid wave function
        for filling factor $\nu=1/6$ of the LLL.
        The values of the anisotropy parameter are 
        $\alpha=0$ (Filled circle),
        $\alpha=4$ (Filled diamond) and
        $\alpha=6$ (Filled square).
        Note that the value $\alpha=0$ represents an 
        isotropic Fermi liquid phase.
        Electrons interact with an anisotropic Coulomb interaction
        potential, $v_{\gamma=2}(x,y)$.
}
\end{figure}
%
%
%
%
%
%
%
%
\begin{figure}[!t]
\includegraphics[width=3.4in]{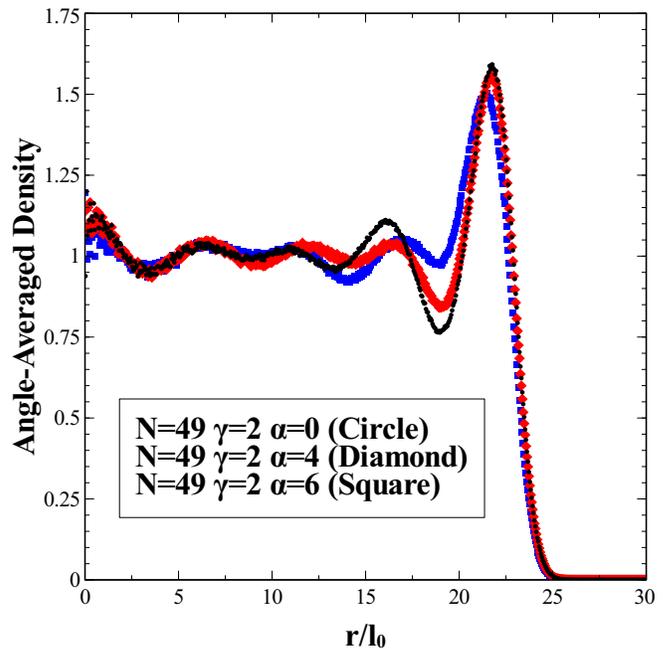}
\caption{  \label{fig:rhon49}
        Angle-averaged density function, $\rho(r)/\rho_{0}$
        as a function of the dimensionless distance, $r/l_0$ 
        from the center of the disk for a system of
        $N=49$ electrons.
        Electrons interact with an anisotropic Coulomb
        interaction potential, $v_{\gamma=2}(x,y)$.
        The system is described by an anisotropic BRS Fermi liquid wave function
        for filling factor $\nu=1/6$ of the LLL.
        The values of the anisotropy parameter are 
        $\alpha=0$ (Filled circle),
        $\alpha=4$ (Filled diamond) and
        $\alpha=6$ (Filled square).
        Note that the value $\alpha=0$ represents an 
        isotropic Fermi liquid phase.
        Note the persistence of an "edge region" of finite
        width at the edge of the disk.
        The quantity, $\rho_0=\nu/(2 \, \pi \, l_0^2)$ represents
        the uniform density value for the system under consideration.
}
\end{figure}
%
%
%
%
%
%
%
%
%
%
%

%
%
%
%
%
There are some qualitative similarities between the $\Delta \epsilon_{\alpha}$ vs $\alpha$
dependence in the current work to an earlier study~\cite{cite2011d}
that dealt with the possibility of an anisotropic quantum Hall liquid state at filling factor $\nu=9/2$.
In that study, we considered finite systems of electrons interacting
with an isotropic (though non-Coulomb) interaction potential that was properly raised in the
second excited LL (the level with quantum number, $n=2$).
In such an instance, we argued for the possibility of liquid crystalline phases where rotational 
symmetry is spontaneously broken (since the interaction potential is isotropic).
We found out that the optimal value, 
$\alpha_0$ 
for which a minimum energy  is obtained 
tends to be size-dependent for the $\nu=9/2$ state.
It was $\alpha_0 \approx 4$ for a system of $N=5$ electrons,
it became $\alpha_0 \approx 3$ for $N=13$  electrons
and eventually stabilized to the value $\alpha_0 \approx 2$
for the largest systems considered ($21 \leq  N \leq 49$).
We attributed the variations of $\alpha_0$ for small systems ($5 \le N \le 13$) 
to the small size and "edge" effects.
However, we argued that the optimal value $\alpha_0=2$ is not casual.
In fact we suggested that the optimal choice, $\alpha_0=2$ for the $\nu=9/2$ state 
is approximately set by the location of the dominant cusp of the $v_{n=2}(r)$ potential 
which represents the effective interaction potential between electrons 
in the upper LL with quantum number $n=2$ that is half-filled~\cite{cite2011d}.
We remind the reader that the interaction potential, $v_{n=2}(r)$ is
isotropic but  non-monotonic (has cusps).
It can be exactly calculated in real space~\cite{orionjltp}.
It differs substantially from an isotropic and/or anisotropic Coulomb interaction potential
(like the one in the current work)
for all excited LLs where $n \geq 1$.

For the case of the anisotropic Coulomb interaction potential, $v_{\gamma=2}(x,y)$
we found out that, except for the smallest of the systems of electrons considered ($N=5$),
the lowest energy of an anisotropic state is achieved for $\alpha_0=5$
in all other cases ($9 \leq N \leq 57$). 
Specifically speaking, the results for the $v_{\gamma=2}(x,y)$ interaction potential
indicate that the optimal value, $\alpha_0$ 
shows some size-dependence for the smallest systems considered, 
but then quickly settles to $\alpha_0 \approx 5$ when the system size increases.
It is hard to tell if this value is hinting at something of significance. 
%
Generally speaking, we would argue that the optimal choice for $\alpha_0$ is set by the 
strength of the interaction anisotropy parameter, $\gamma$.
It is very tempting to say that the optimal value, $\alpha_0$
is set by the ratio $v_{\gamma}(x,y=0)/v_{\gamma}(x=0,y) \propto \gamma^2$ for $|x| \approx |y|$
but this appears to be too crude.
Nevertheless, this statement might still bear some truth since simulations with $N=25$ electrons 
indicate that the optimal choice, $\alpha_0$ decreases when $\gamma$ decreases.
For instance, our simulation results for $N=25$ electrons indicate that
the optimal value is $\alpha_0=4$ for $v_{\gamma=1.25}(x,y)$ and 
$v_{\gamma=1.50}(x,y)$ interaction potentials.
However, as can also be seen from the results in Fig.~\ref{fig:ener_n25_g}, 
the value $\alpha_0$ 
increases  to $\alpha_0=5$ for $v_{\gamma=1.75}(x,y)$, $v_{\gamma=2.00}(x,y)$,
 and  $v_{\gamma=2.25.}(x,y)$ interaction potentials.
As previously stated, we use only values, $\alpha=0, 1, \ldots, 8$ in our calculations.
%
%
%
%
%
%
%
In Fig.~\ref{fig:gn57} we plot the angle-averaged pair distribution function, 
$g(r)=\int_{0}^{2 \, \pi} \frac{d \, \theta}{2 \, \pi} \, g(r,\theta)$ 
for $\alpha=0, 4$ and $6$ corresponding to a system of $N=57$ electrons,
the largest that we were able to consider.
One notices that the major peak of $g(r)$ becomes less pronounced and 
very slightly shifts to larger values of $r$ as $\alpha$ increases. 
The short-range behavior of the $g(r)$ also changes as $\alpha$ increases.
We found that $g(r)$ changes from approximately  
$\propto r^{14}$ ($\alpha=0$) to approximately $\propto r^{10}$ ($\alpha \neq 0$).
Such a shift  is expected from general considerations of the nature of the 
magnitude square of the wave function at small separation distances 
between electrons.
%
%
The differences in the $g(r)$-s may also be the result of 
different correlations in different directions canceling.
%
%
%
%
%
%
%
%
%
%
%
%
In Fig.~\ref{fig:rhon49} we plot the angle-averaged density function, 
$\rho(r)=\int_{0}^{2 \, \pi} \frac{d \, \theta}{2 \, \pi} \, \rho(r,\theta)$
relative to its uniform density value, $\rho_0$ as a function of
the dimensionless distance parameter, $r/l_0$
for several BRS wave functions with different values of the anisotropy parameter, $\alpha$.
The density function seems to be sensitive to the wave function 
parameters in the central region of the disk, but less at the edge.
This might be an indication of interesting topological 
quantum Hall edge effects~\cite{ijmpb2012}
where anisotropy might play some subtle role.

%
%
%
%
%
%
%
%
Another way to incorporate the effects of anisotropy 
into a typical quantum Hall wave function is
through appropriate modifications of the CF theory.
For instance, one can write a family of microscopic CF wave functions 
that incorporate electron mass anisotropy
by a rescaling of the electron coordinates~\cite{balram}.
In this model one starts with electrons that have
anisotropic mass ($m_x \neq m_y$) and 
then sets up the single-particle particle states in
terms of complex coordinates that depend on a 
mass anisotropy parameter
(with $x$ and $y$ coordinates scaled differently).
The resulting CF wave functionn is anisotropic since in this 
approach the single-particle states are anisotropic by construction.
The problem of electrons with anisotropic mass
interacting with an isotropic Coulomb interaction potential is 
effectively equivalent to the problem of electrons with
isotropic mass interacting with an anisotropic Coulomb 
interaction potential as the one considered in our work.
Thus, we are basically dealing with the some problem 
despite apparent differences in formalism, namely, 
the interaction anisotropy parameter $\gamma$ of the $v_{\gamma}(x,y)$ potential 
may be interpreted as a mass anisotropy parameter
in the anisotropic CF approach of systems of electrons interacting via the usual
isotropic Coulomb interaction potential.
At this juncture, apart from the fact that we are looking at different filling factors,  
it is important to remark a striking difference between the wave function in Eq.(\ref{bosealpha}) 
and  the anisotropic CF wave functions already mentioned~\cite{balram}.
Anisotropy in our case is introduced at the two-body level 
of the wave function while anisotropy originates from single-particle states
within the anisotropic CF wave function framework.

It is interesting to point out that
small-$N$ results for anisotropic CF states at Laughlin filling factor $\nu=1/3$ 
seem to suggest that energy is very insensitive to
anisotropy for up to fairly large values of the effective mass/interaction
anisotropy parameter~\cite{balram}.
Our calculations for the Fermi liquid state at filling factor $\nu=1/6$ 
pursue a different avenue, but seem to indicate that the energy
of the $\nu=1/6$ state is quite sensitive to the increasing anisotropy 
at least for the specific choice of the wave function that we made.
Absence of an energy gap for the $\nu=1/6$ Fermi liquid state 
under consideration might suggest that such a state is easily 
deformable and, thus, might be prone to destabilization.
If there is some truth in this statement, a more general hint comes 
from this behavior, namely, all the other Fermi liquid states in the LLL might
show similar tendencies since all Fermi liquid  states at filling factors of the form 
$\nu=1/2$, $1/4$ and $1/6$ share similar characteristics.
Thus, one might be tempted to say that all even-denominator filled Fermi liquid states
in the LLL (including the ones at $\nu=1/2$ and $\nu=1/4$)
might be prone to destabilization by an anisotropic Coulomb interaction 
of the form considered here through the same mechanism.

%
%
%
%
To conclude, 
in this work we introduced the idea that the impact of 
a weak anisotropic electron-electron interaction perturbation 
possibly originating from the piezoelectric property of the GaAs substrate 
may stabilize novel electronic liquid crystalline phases even in the LLL. 
%
%
In particular, we envisioned the possibility that the very fragile 
isotropic Fermi liquid state at filling factor $\nu=1/6$ of the LLL 
can be destabilized by an electron-electron anisotropic perturbation. 
%
Our calculations suggest that this happens for the case of a phenomenological
anisotropic Coulomb interaction potential, $v_{\gamma}(x,y)$ at all values
of the interaction anisotropy parameter we considered ($1.25 \leq \gamma \leq 2.25$)
for all the sizes of the finite systems of electrons that we considered ($5 \leq N \leq 57$).
Our finite QMC results suggest the stabilization of a novel anisotropic liquid crystalline 
phase of electrons due to anisotropy whose experimental signatures may be detectable at ultra-low 
temperatures of order $1 \, mK$. 
It is conjectured that the same tendency toward liquid crystalline order 
in presence of an anisotropic interaction potentional may be observed
at other known even-denominator filled Fermi liquid states in the LLL 
(for instance, $\nu=1/2$ and $1/4$).

\vspace{0.5cm}

\acknowledgments

This research was supported in part by U.S. Army Research Office (ARO) Grant No. W911NF-13-1-0139
and National Science Foundation (NSF) Grant No. DMR-1410350.



\end{document}